\documentclass[journal]{IEEEtran}
\IEEEoverridecommandlockouts
\usepackage{cite}
\usepackage{amsmath,amssymb,amsfonts}
\usepackage{algorithmic}
\usepackage{graphicx}
\usepackage{textcomp}
\usepackage{xcolor}
\usepackage{comment}

\ifCLASSINFOpdf
\else

\fi

\begin{document}
\title{ECE496Y Final Report: Edge Machine Learning for Detecting Freezing of Gait in Parkinson's Patients}
\author{Purnoor Ghuman,
        Tyama Lyall,
        Usama Mahboob, 
        Alia Aamir\\
        Supervisor: Professor~Xilin~Liu\\ The Edward S. Rogers Sr. Department of Electrical and Computer Engineering \\University of Toronto
}

\markboth{ECE496Y Design Project Course Final Report, Winter 2022}
{Shell \MakeLowercase{\textit{et al.}}: Bare Demo of IEEEtran.cls for IEEE Journals}

\maketitle

\begin{abstract}
Parkinson's disease is a common neurological disease, entailing a multitude of motor deficiency symptoms. In this project, we developed a device with an uploaded edge machine learning algorithm that can detect the onset of freezing of gait symptoms in a Parkinson's patient. The algorithm achieved an accuracy of 83.7\% in a validation using data from ten patients. The model was deployed in a microcontroller Arduino Nano 33 BLE Sense Board model and validated in real-time operation with data streamed to the microcontroller from a computer. 
\end{abstract}


\IEEEpeerreviewmaketitle

\section{Introduction}

\IEEEPARstart{P}{arkinson's} disease (PD) is a neurodegenerative disorder, the second most common following after Alzheimer’s disease. PD patients experience a multitude of movement disorders, in particular, one of the common movement disorders is the Freezing of Gait (FoG), prevalent in the late stages of PD \cite{bikias2021deepfog}. A patient’s walking and turning is impeded, thereby increasing instability and the risk of falls, and as a result potential injuries.

Previous treatment for FoG has included a variety of options; such as medical treatments and invasive procedures, each with varying results and potential adverse effects \cite{su2020gait}. Studies have also been done on cognitive exercises such as Rhythmic Auditory Stimulation (RAS). RAS introduces an external auditory stimulus for patients’ to synchronize their steps, enabling them to initiate walking with a higher rate of success. This solution requires PD patients to be aided by medical professionals, but there is yet to be a feasible solution that detects the onset of FoG to make this form of treatment feasible \cite{liu2016design}. Innovative research has gone into wearable devices with acceleration sensors that detect the onset of FoG via the frequency components of limbs, but had inconsistent results due to inability to detect FoG with higher accuracy \cite{bachlin2009wearable}.

The above studies give way to our motivation, which is the utilization of machine learning. Machine learning algorithms can be trained on a set of patient data and with some optimization, learn the onset of FoG episodes. This allows the timely delivery of rhythmic auditory and electrical stimulation to the patient to enable them to overcome difficulty with walking initiation and turning [4]. The motivation for this project is to implement a non-invasive, feasible method for detecting the onset of FoG in PD patients. The gap is a device which can run a machine learning algorithm locally on an embedded system. This device should accurately predict, based on input data of the patient’s movements, whether a patient is experiencing FoG or not and output a signal indicating this to an external source. An algorithm like this which processes data locally is called an edge machine learning algorithm, resulting in more convenience for PD patients wearing it to successfully detect the onset of FoG symptoms \cite{merenda2020edge,borzi2021prediction,liu2021edge,liu2021system}.

The goal of this project is to develop a microcontroller device with an uploaded edge machine learning algorithm that can detect the onset of freezing of gait symptoms in a PD patient from their acceleration data in a timely manner with limited power resources \cite{liu2014low,liu2021edge}.

\begin{figure}[ht]
\centering
\includegraphics[width=.9\linewidth]{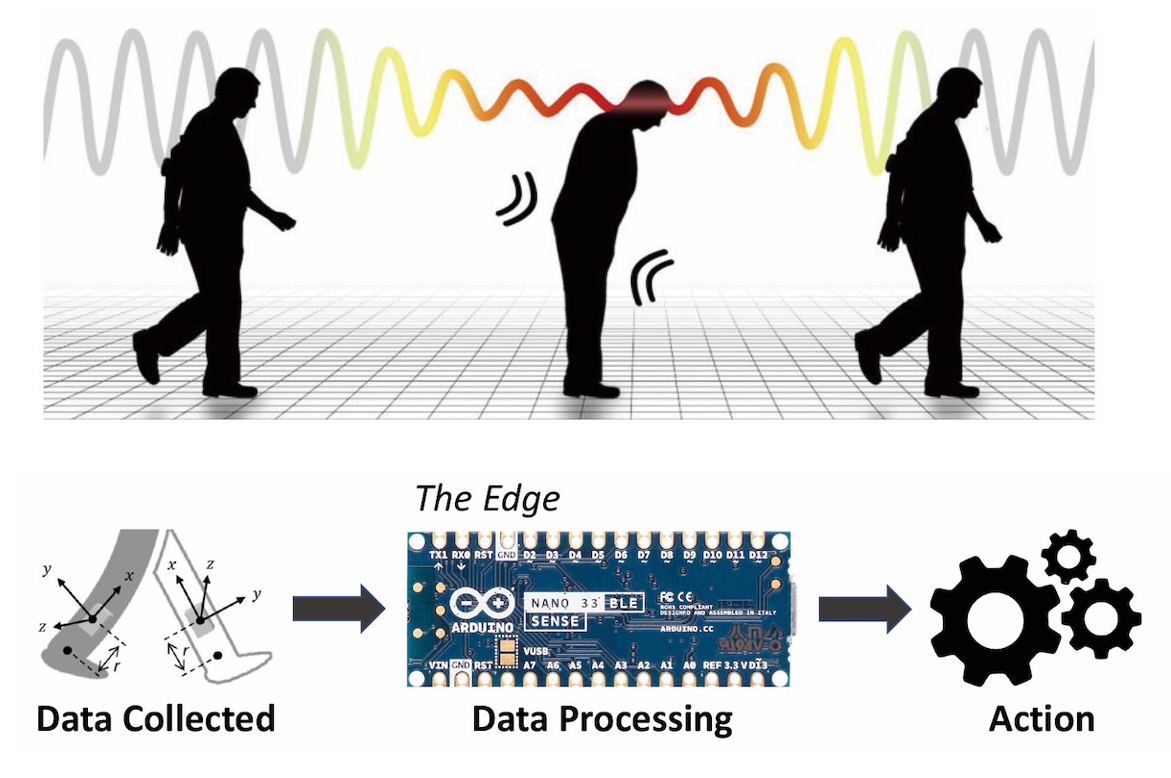}
\caption{Introduction of proposed FoG detection using edge machine learning.}
\label{fig:intro}
\end{figure}

\section{Methods}
\subsection{System Level Overview}

\begin{figure}[ht]
\centering
\includegraphics[width=1\linewidth]{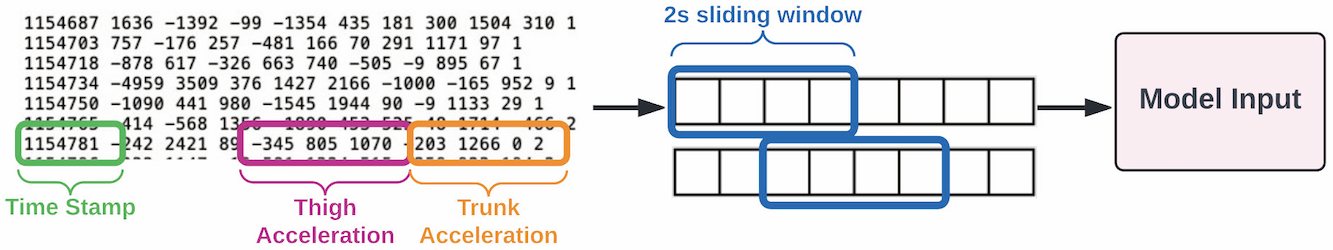}
\caption{Illustration of the data processing steps.}
\label{fig:data_processing}
\end{figure}

The final design being implemented by the team consists of two main aspects. The first part is the training and development of the machine learning model. For this part, the initial step is the cleaning of the PD patient acceleration data, shown in Figure \ref{fig:data_processing}. Unwanted timestamps, where the patient was not walking, are removed in this step. Also, only thigh acceleration is considered in this project because an adequate idea of a patient’s gait can be determined from their thigh acceleration data.

\begin{figure}[!ht]
\centering
\includegraphics[width=1\linewidth]{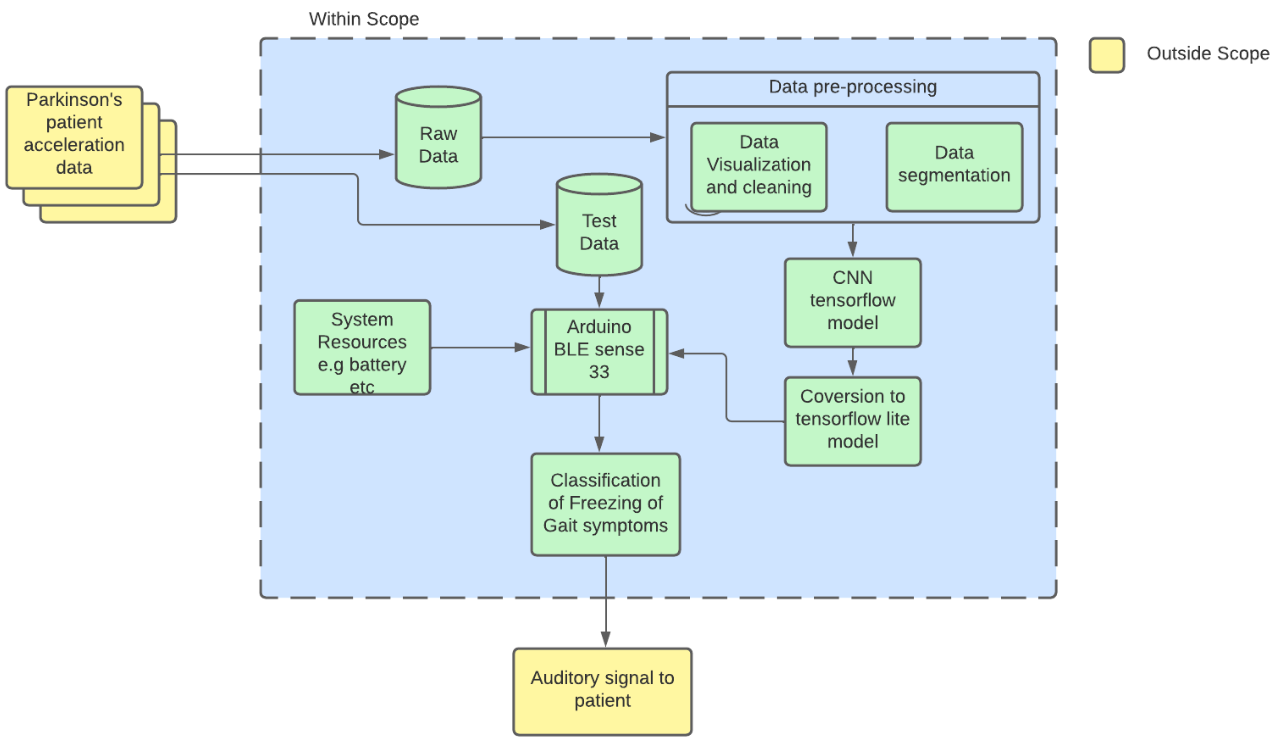}
\caption{Simplified system context diagram.}
\label{fig:sys_diagram}
\end{figure}

The cleaned data, consisting of multiple timestamps with their corresponding acceleration values and classifications, is then segmented into windows of 2 seconds, with each window having its own classification as FoG, denoted as a 1, or No-FoG, denoted as a 2. Next, our team trained a Convolutional Neural Network (CNN) with the time windows as inputs to classify each 2 second window as FoG or No-FoG.

In the second part of the project, the trained CNN model is then extracted as a TensorFlow Lite model and uploaded onto an Arduino BLE Sense 33 microcontroller \cite{liu2021edge}. The input raw test data is stored on a computer in a CSV file and written into the board using a script which utilizes the “pandas” library. A connection between the script and the Arduino port is made to write data into the board, process a window using the loaded CNN model, and extract the result, all done in real time.

The system context diagram shown in Figure \ref{fig:sys_diagram} summarizes the above information succinctly, showing each working module of the project.

\subsection{Module Level Design}

\subsubsection{Data Pre-processing}
We took the data from the .csv files and separated them in terms of the type of acceleration data e.g., thigh vs. ankle acceleration. Used the matplot library to plot graphs of acceleration values against a time axis. Analyzed the graphs to determine where and how often the data is useful as there are segments in time where the patient is not moving so some time intervals are not useful for our classification task.

We converted patient data files to suitable format (.csv) for later visualization and manipulation. Removed data not a part of the experiment (data with annotation ‘0’), determined by reading through provided documentation. In addition, from data visualization, removed periods of no-walking for each of the csv files using Microsoft Excel. 

The results of data cleanup was data ready for further manipulation and usage by model. Data visualization enabled further data cleanup through graphing the data points and analyzing them visually. 

\subsubsection{Data Segmentation and Augmentation}

We developed a subroutine to divide the continuous acceleration data into two second windows and return a list of data points, where each element represents one 2 second window, with the corresponding accelerations across it and a corresponding list of labels for each window. From the patient data, FoG is denoted with a 2 and No-Fog is denoted with a 1, but the labels are passed through a function to change the labels to 0 for No-FoG and 1 for FoG to correspond with labels typically used in binary classification tasks. To augment the data and yield more data points, a sliding window approach was used, where each window has an overlap with the previous of one second.

In order to generate more instances of the minority class, instances of FoG in our case, oversampling was performed. This involves randomly selecting a predetermined amount of data points from the minority class to duplicate in order to improve the model’s ability to learn on this class \cite{brownlee2020random}.

The final results of this module is an augmented dataset, ready to be passed into a model. The dimension of the input data is (8711, 129, 3), where 8711 is the number of two second windows created, 129 is the number of timestamps in each window, and 3 is the number of accelerations provided for each timestamp, one for each dimension x, y, z.

\begin{figure}[!ht]
\centering
\includegraphics[width=1\linewidth]{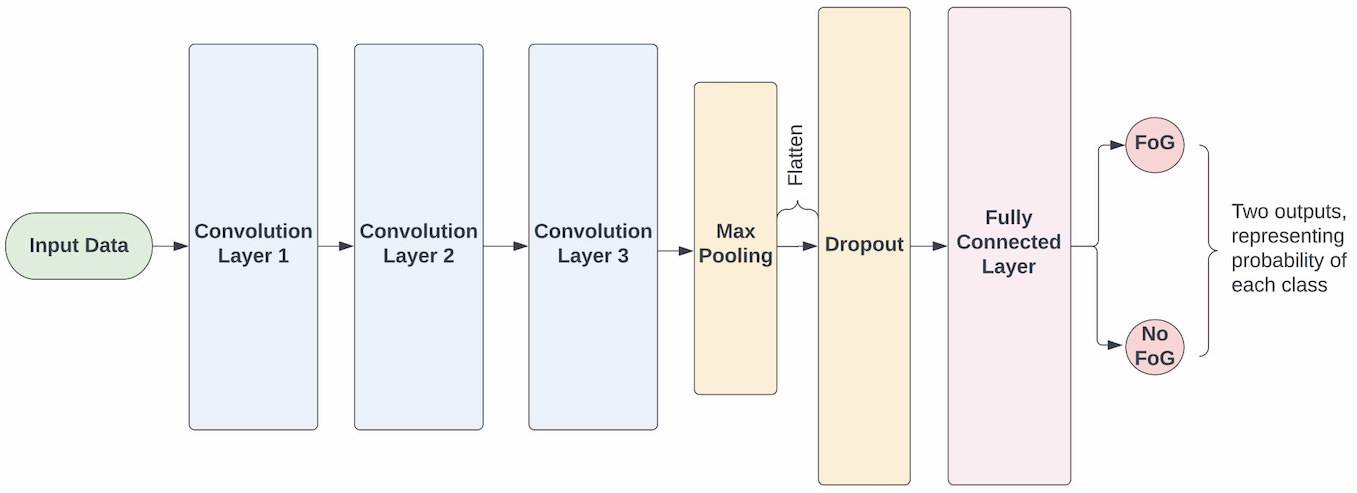}
\caption{A simplified model diagram.}
\label{fig:model}
\end{figure}

\subsubsection{Model Implementation}

We model design entailed the initial research and outlining of initial model implementation. Decisions were made here to start building CNN with a simple 2-layer MaxPooling and Convolutional layers model. We also made the decision to use an optimizer to adjust weights and the learning rate to reduce losses during training \cite{doshi2019various}. Concluded using Adaptive Moment Estimation (Adam) as an optimizer during model training, due to its fast convergence and efficient process \cite{doshi2019various,liu2020fully}.

We researched and utilized Keras library for the implementation of a CNN model [6, 14]. From the initial model design, obtained additional information regarding how to structure the model. Initial implementation used in training included the usage of 2 Convolution layers, a MaxPooling layer, and 2 Dense layers \cite{allibhai2018building}. Dense layers were utilized, in which we defined an activation function, to change dimensions of expanded units into a classification output \cite{sharma2020keras,liu2021energy}. The model was then compiled using the previously mentioned ADAM optimizer and our choice of loss function \cite{doshi2019various}. The loss function we used was binary cross entropy loss, utilized for the purpose of determining how well our predictions are \cite{sharma2017activation,liu2021system}.

The model was trained and validated using the dataset splits previously mentioned, along with the usage of the Keras library fit method. This method enables model training and validation, with the corresponding ability to adjust several parameters [17].

The final results of this module were our initial model, which can be found in Appendix F. We also conducted and received results of its initial training and validation, which can be found in Appendix G. From observing the validation, as well as the accuracy corresponding to No-FoG and FoG classes, it was determined further iteration was required, which is discussed in the subsequent module.

\subsubsection{Model Optimization}

Selected model hyperparameters to tune (number of filters in convolutional layer, learning rate, epochs, batch size) and used Keras.tuner library and scikitlearn.GridSearchCV to learn optimal values for each one. See Appendix H for values over which tuning was conducted on each hyperparameter and corresponding code \cite{lau2019walkthrough}.

Regularization techniques are used to prevent the model from overfitting to the training data. Techniques used include the introduction of a dropout layer in the model structure, which randomly drops some inputs to improve the model’s generalization [18] and the use of early stopping during training, which will stop training if the validation loss stops decreasing over many epochs. In order to improve the model’s accuracy on the minority class, class weights were calculated based on the ratio of positive to negative examples in the training data and those weights are used to bias the classifications made during training.

The final result was a model with hyperparameters tuned using the validation set to maximize accuracy on both classes and to generalize well to brand new data.

\subsection{Deployment to Microcontroller}

We worked on converting tensorflow model to a tensorflow lite model and optimized it for memory footprint in order to ensure a small size so it can fit the memory constraint \cite{liu2017fully}. In order to optimize we decided to analyze the final model and merge any common nodes and remove any unused nodes from the graph. We also decided to freeze the graph by converting the weights into constants so it would take up less space than having them as variables. We also decided to quantize the inputs and outputs by decreasing the precision of the variables so that they would take up less space. This only resulted in a minimal decrease in accuracy so we decided to keep it.

We set up the Arduino board to upload the model and run continuously while taking in inputs and calculating output of the model and displaying it on the serial monitor. We had to set up error logging to ensure any errors caught during inference are caught and had to account for how much dynamic and static memory is available on the board to determine what size of the input can be sent to the board at a time to perform inference and that there was enough space for the board to store the resulting output classification.

We designed an efficient method to write raw acceleration data into the microcontroller, which would be then processed by the embedded machine learning model (Module 2.3.5), and read the produced output from the board. This is implemented using a Python script, which utilizes the pandas library, to write data “chunks”, or windows, into the Arduino board [19]. After the script has written enough data to cover a full window of patient accelerations, the window is passed into the model for processing, which is then read by the script, as shown in Appendix L.

Since an Arduino board cannot take multiple rows of input at once, all data needs to be passed into the board individually, byte-by-byte. This presents a challenge, as on the Arduino board side, we need to know when to stop collecting data from the Python script and process. This is done by defining two states called ‘Read’ and ‘Process’. In the ‘Read’ state, the board reads the data provided by a computer and populates the appropriate variables, until the right number of variables have been populated. At that point, the state switches to ‘Process’ and data is passed into the machine learning model, after which an output is produced and the state is switched back to ‘Read’.

The implementation of this module results in our raw test data being stored on a computer and streamed directly to the Arduino board, instead of taking up space on the board itself.

\section{Results}

\subsection{Assessment of Final Design}

Overall, assessment of the final design can be broken down into two primary parts: the binary classification model and the hardware implementation. While the two aspects are interconnected and work together, their performance can be assessed separately.

The model has been tuned using a validation set and trained on the training set, and performs reasonably well on new data, with an overall accuracy of 83.7\%. The accuracy on no FoG examples was 84\% while the accuracy on FoG examples was 80\%. The accuracy is not as high as the desired threshold originally set by the team, largely due to the imbalance in the data provided. Development and training of the model was hindered by the fact that the dataset contained much more data for No-FoG than data featuring instances of FoG, making it challenging to develop a model that could recognize FoG instances with high accuracy. The original plan to use an SVM for classification was discarded, as it was impossible to train an SVM that was capable of differentiating between the two classes; it would classify every point either as one class or the other. The team moved on to researching deep learning approaches and eventually decided on a CNN due to their weight sharing capabilities, which lower the number of learned parameters and therefore the size of the model. Due to space constraints, the final model has three convolutional layers followed by one fully connected layer with two outputs, one for each class. Through data standardization, weight biasing, and various regularization techniques outlined in section 2.3.4, a final model design was decided on. The final model size was ~0.4 MB, which is well below the memory requirement established by the team.

The hardware implementation involves downloading the trained CNN model to a header file which can be uploaded onto the Arduino microcontroller. From there, input data is written to the board using a Python script. Another program is also written to the board, which can take in the data being written and segment it continuously into two second windows. Each window is then passed into the trained model for classification and the output of the model will be written back to the computer to indicate when FoG is detected. Further details on how each requirement is tested and verified by the above design is discussed in the next section.

\subsubsection{Verification of Correct Input into Arduino Board}

To confirm that the Arduino board is correctly receiving the acceleration data being provided from a CSV File, a simple check is performed in the Python script providing the data. In this check, one acceleration value is provided as input into the board. The board then takes that acceleration value and increments it by one and prints it on the monitor. The Python script, after waiting for something to be printed on the Arduino terminal, consumes the printed value and prints it on its own terminal. The input value and the output received from the Arduino board is shown in Figure M3 in Appendix M. The Python code performing this check can be seen in Figure M1 in Appendix M and the Arduino code is shown in Figure M2 in Appendix M. This check confirms that acceleration data being read into the board is received correctly and that the processing done locally is also correct, as the output value must always be an increment by one of the input value.

\subsubsection{Verification of Correct Input into Arduino Board}

To confirm that the machine learning model is, in fact, uploaded on the board, we must place the “model.h” file containing the model in the same folder as the Arduino file uploading it on the board. After the upload happens, it must be ensured that the upload happened successfully and that the correct number of bytes are written on to the board (~400,000 in this case). This can be confirmed by looking at the output of the upload shown in Figure 3, where a successful upload is shown and 384952 bytes of memory of the board has been filled.

\subsubsection{Verification of ML Model Accuracy}
\begin{figure}[ht]
\centering
\includegraphics[width=1\linewidth]{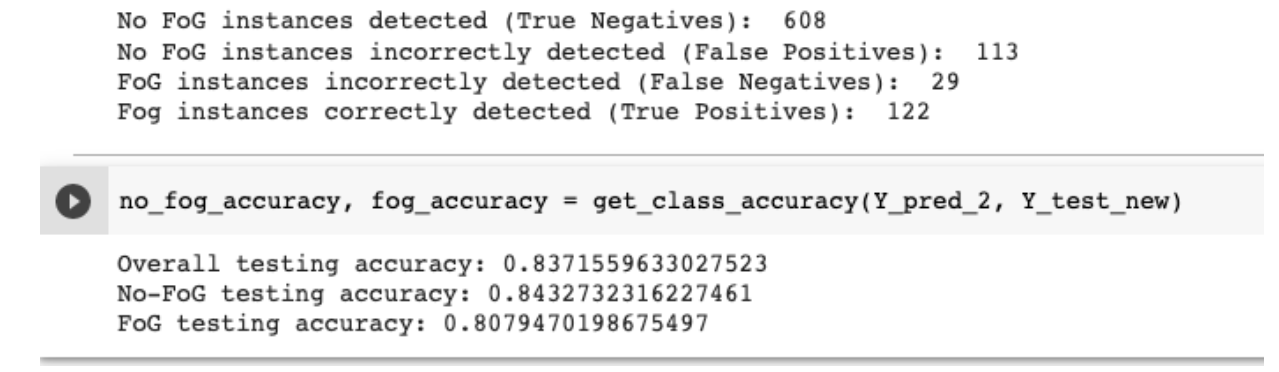}
\includegraphics[width=.7\linewidth]{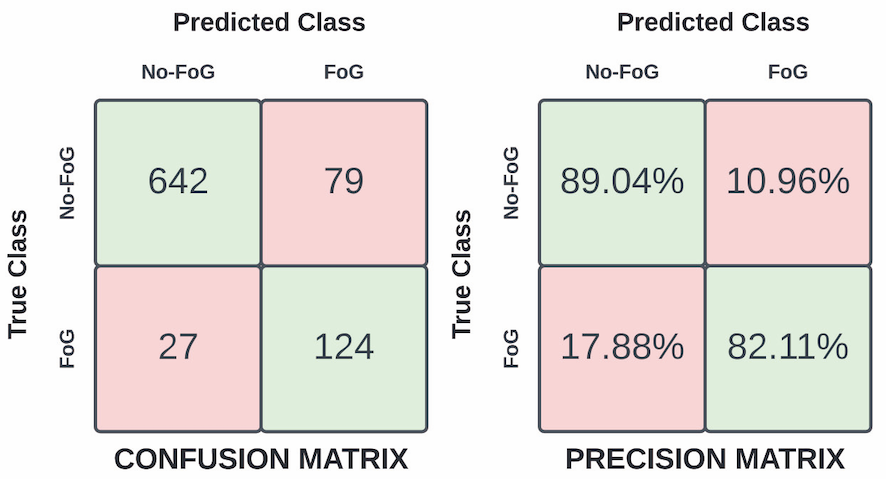}
\caption{Confusion matrix metrics and accuracy on each class for final model.}
\label{fig:confusion}
\end{figure}

To confirm that the model’s accuracy meets our objective, we generated the trained model’s predictions on the test set, which it had never seen before. Then, compare the predictions with the true labels and generate a percentage from that which represents the accuracy.The accuracy on no FoG examples was 84\% and the accuracy on FoG examples was 80\%, yielding an overall accuracy of 83.7\%. To further quantify the model’s performance, we can calculate its accuracy on each class, and generate a confusion matrix, which will give us the number of false positives and negatives as seen in Fig. \ref{fig:confusion}.

\subsubsection{Verification of Model Memory Requirement}

To confirm that the model size will not exceed the 1MB flash memory capability of the microcontroller, we can write code to generate a file model.h containing our trained model, and then output the size of that file in bytes. As seen in Figure 5, the model size is 298,600 bytes, or around 0.3 MB, meaning it meets the memory requirement.

\subsubsection{Verification of Classification Output}
\begin{figure}[!ht]
\centering
\includegraphics[width=1\linewidth]{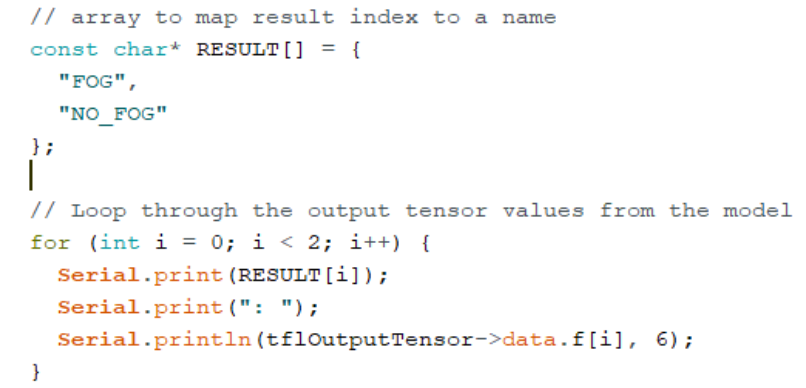}
\caption{Code to print classification onto serial monitor.}
\label{fig:code}
\end{figure}

For classification, a 2 second window (input) is sent to the board which uses the uploaded model to perform inference based on this input and outputs a classification result of either FoG or No-FoG. This result is then displayed on the serial monitor of the computer that is connected to the board as the board itself has no screen to display the result. As it has not been fully implemented on the board yet the inference and output will be shown as follows in Fig. \ref{fig:code}.

\subsubsection{Verification of Memory Resources Utilized on the Board}
Given that the board has a total of 1 megabyte of flash memory, we had to make sure that the size of our classification sketch was below that. This led to a limitation on the number of layers in the model, otherwise the size would increase exponentially. We had to find a good balance between having a complex model with multiple layers and weights to ensure good accuracy but also making sure that its size does not exceed the 1 megabyte constraint.

The model uses 384952 bytes of flash and only 21\% of the dynamic memory and 40\% of storage memory. The rest will be used to store input (2 second windows) to be sent to the model for inference. Therefore the memory constraint set by the board’s onboard memory is satisfied.

\section{Conclusion}

After extensive research and multiple iterations, the team was successful in producing a CNN machine learning model that achieves the accuracy and specificity set out in the requirements. We were able to meet all the requirements laid out in our project and therefore accomplish our goal of correctly predicting the onset of FoG based on the acceleration data provided in real time that is run locally on the microcontroller.

As can be seen in the verification and testing section, the model is small enough to be uploaded on the board’s limited memory and is able to run inference within 5 seconds. Therefore, it is suitable to help patients identify FoG onset while giving them enough time to be able to react such as sitting down before it becomes active. The accuracy of the model is high enough to be utilized by patients without giving inaccurate results.

Throughout the project, we had to make multiple changes in order to achieve an accurate model. For example, initially we had planned to extract features from the data manually, but that yielded very low accuracy \cite{liu201512}. We had to iterate over multiple variations of the model, from using an SVM to a Neural Network (NN) to finally using an CNN. Through this project we realized that in order for the model to be compatible with the microcontroller we were restricted to a few libraries such as Tensorflow. From this project, we learned that there are many different algorithms that can be used to classify FoG onset, but only a few finely tuned models were able to achieve all the requirements such as a high accuracy and a small enough memory footprint so that it can be uploaded to the microcontroller.

In the future, we can potentially use the acceleration data taken directly from the microcontroller’s sensors as a patient walks to run inference on data as it is collected to be used in a real-case scenario \cite{liu2013low}.

\bibliographystyle{IEEEtran}
\bibliography{ref}

\ifCLASSOPTIONcaptionsoff
  \newpage
\fi

\end{document}